%% file: main.tex
\documentclass[conference]{IEEEtran}
\IEEEoverridecommandlockouts
\usepackage{cite}
\usepackage{amsmath,amssymb,amsfonts}
\usepackage{fancyhdr}
\usepackage[ruled]{algorithm2e}
\usepackage{graphicx}
\usepackage{textcomp}
\usepackage{xcolor}
\usepackage{url}
\def\BibTeX{{\rm B\kern-.05em{\sc i\kern-.025em b}\kern-.08em
    T\kern-.1667em\lower.7ex\hbox{E}\kern-.125emX}}
\begin{document}
\pagestyle{fancy}
\fancyhead[L]{Accepted as WIP in MLCAD 2024}
\title{VLSI Hypergraph Partitioning with Deep Learning}

\author{\IEEEauthorblockN{Muhammad Hadir Khan, 
Bugra Onal, 
Eren Dogan,
Matthew R. Guthaus }\\
\IEEEauthorblockA{Computer Science and Engineering, University of California Santa Cruz, Santa Cruz, CA 95064\\ 
\{mkhan33, bonal, erdogan, mrg\}@ucsc.edu}}
% \author{Do not distribute. For anonymous review only.}

\newcommand{\fixme}[1]{{\bf FIXME: {#1}}}

\input{variables}

\maketitle

\begin{abstract}
Partitioning is a known problem in computer science and is critical in chip design workflows, as advancements in this area can significantly influence design quality and efficiency. Deep Learning (DL) techniques, particularly those involving Graph Neural Networks (GNNs), have demonstrated strong performance in various node, edge, and graph prediction tasks using both inductive and transductive learning methods. A notable area of recent interest within GNNs are pooling layers and their application to graph partitioning. While these methods have yielded promising results across social, computational, and other random graphs, their effectiveness has not yet been explored in the context of VLSI hypergraph netlists. In this study, we introduce a new set of synthetic partitioning benchmarks that emulate real-world netlist characteristics and possess a known upper bound for solution cut quality. We distinguish these benchmarks with the prior work and evaluate existing state-of-the-art partitioning algorithms alongside GNN-based approaches, highlighting their respective advantages and disadvantages.
\end{abstract}

%\begin{IEEEkeywords}
%component, formatting, style, styling, insert
%\end{IEEEkeywords}

\input{intro}

\input{background}

\input{implementation}

\input{methodology.tex}
\input{results}
\input{conclusion}

\clearpage
\newpage
{\small
\bibliographystyle{ieeetr}
\bibliography{main}
}
\end{document}

%% file: variables.tex
% These can be used in the paper by using \ourName for example.

\def \ourName {PERRDI\xspace} 
\def \ourNameLong {Partitioning Examples with Rent's Rule Derived Information (PERRDI)\xspace}
\def \erdos {Erd\H{o}s-Reny\'i\xspace}

%% file: intro.tex
\section{Introduction}

%Partitioning uses in EDA flow - placement, hierarchical abstraction, emulation systems
Modern Electronic Design and Automation (EDA) tools use partitioning throughout the design flow. This may be during placement, simulation, emulation, or any number of other steps. Improving partitioners can directly make these steps work better and improve design quality of results.  

%State-of-the-art partitioners - FM, KL, Spectral, multilevel
Fundamental partitioning heuristics like FM~\cite{fm} and KL~\cite{kl} work reasonably well for small graphs, but not for large ones. Spectral partitioning~\cite{spectral} improved upon this by using a graph's topological structure through its Eigenvectors to induce an ordering used for partitioning. This spectral ordering was frequently used to initialize the partitions and then the fundamental heuristics like FM were used for refinement. The state-of-the-art partitioning algorithms, however, use multi-level~\cite{mlpart, hmetis} approaches that cluster and refine so that fundamental algorithms can scale to large graphs.

% ML Partitioning
Recently, partitioning has gained interest in the Machine Learning (ML) community for pooling in Graph Neural Networks (GNNs)~\cite{gnn_survey}. Pooling performs clustering on a graph to create a reduced graph with a representative set of features on the fewer, new graph nodes. There have been numerous algorithms proposed for pooling in GNNs~\cite{asymcheegercutpool, mincutpool, dmonpool, diffpool}. 

% What is GAP and why we use GAP?
Similar to pooling, the Generalizable Approximate Graph Partitioning Framework (GAP)~\cite{gap-old} performs partitioning on general graphs by relying on features generated from the Principal Component Analysis (PCA) components. GAP is similar to spectral partitioning techniques~\cite{alpert1995spectral} except that a linear ordering is avoided and instead an ML classification with a trainable inference step is utilized. Compared to the other ML pooling techiques, GAP is the only work to analyze inductive learning on unseen graphs. 

% Our main motivation: VLSI designs
The new ML partitioning and pooling algorithms, however, have not been considered for use in the VLSI domain. These pooling techniques are instead evaluated on graph classification benchmarks~\cite{KKMMN2016}, node classification datasets (Cora, Citeseer, and Pubmed), computation graphs~\cite{gap-old}, or synthetic random graphs~\cite{gap-old}. Graphs from VLSI design problems have unique characteristics compared to these, however. Most importantly, VLSI designs are hypergraphs and need to be expanded to a simple graph in order to be compatible with most GNN convolution and pooling layers. In particular, VLSI design graphs have gates with limited (e.g., 2-5) inputs and typically a single output. These gates have different areas depending on the number of transistors, their sizes, and their internal connectivity. Fan-out of these gates is also limited (e.g., 3-8) by wire resistance and load capacitance. Some special nets, like clocks and resets, may have high-fan-out that are later buffered. The nets that connect the fan-out are best represented as hyperedges with a single driver and multiple fan-out gates. Input/output pins, or terminal pins, usually have a fixed location and are permanently assigned to a single partition. Most applications of partitioning in VLSI also have a constraint on partition balance so that the gates fit in a given area. Lastly, VLSI design has many objectives such as criticality, wire length, noise, etc. for which net cut is merely a proxy.

% Refer to Rent's rule which talks about avg number of pins per module, BEKU didn't consider this.
In this paper, we propose a new set of synthetic benchmarks, called \ourNameLong, to analyze the advantages and disadvantages of ML partitioning for VLSI design. Our synthetic partitioning benchmarks have a known upper cut bound and resemble typical VLSI design netlist characteristics. We then use these to analyze the best methods to train ML partitioning, examine the run-time, and compare how sub-optimal it is to other partitioners and known upper bound solutions.

%% file: background.tex
\section{Background}

\subsection{Synthetic Partitioning Benchmarks}

% BEKU/MEKU
One prior work, called Bipartitioning Examples with Known Upper bounds (BEKU) and Multi-way Partitioning Examples with Known Upper bounds (MEKU), examined partitioning on VLSI graphs by generating a set of ``known upper bound" synthetic graphs~\cite{beku}. BEKU and MEKU were used to study the optimality of existing serial (and non-ML) partitioners, including MLPart\footnote{Note that ML here stands for Multi-Level not Machine Learning.} and hMETIS.

We've found that the BEKU/MEKU benchmarks are no longer available and the authors have been unable to retrieve archives~\cite{cong}. In addition, from the description in the paper, the BEKU/MEKU benchmarks did not consider many common characteristics of VLSI designs. While they did consider the distribution of different fanout sizes through a Net Distribution Vector (NDV), they did not consider other VLSI design criteria such as gate fan-in and fan-out, fixed terminals, gate sizes, etc. Furthermore, the NDV considered was an absolute number of nets of each size. They did not propose a way to scale the benchmark size. In addition, the cut size was a pre-assigned value that did not relate to realistic designs.

On the other hand, Rent's rule~\cite{rent} is a well known property of hierarchical electronic designs that can be utilized to guide realistic design properties. Rent's rule~\cite{rent} relates the number of external connections $T$ to the number of gates in a design $g$ using two parameters, $t$ and $p$ are the Rent parameters. The parameter $t$ represents the average number of pins per component which ranges from 2-8 and $p$ is the rent exponent which typically ranges from 0.5-0.8. We utilize Rent's rule and gate fan-in/fan-out criteria to improve upon BEKU/MEKU.

\subsection{GNNs}

% (Talk about what GNNs are, and how they are being used in EDA applications)
Graph Neural Networks (GNNs) are types of neural network (NN) models. They input a non-uniform graph, unlike Convolutional Neural Networks (CNNs), which input a regular matrix~\cite{gcn,gat,graphsage}. GNNs can be seen as specialized techniques of CNNs. GNN models can be trained for node prediction, edge prediction, and graph classification tasks. 

% Inductive vs transductive GNNs
GNNs can be trained for either transductive or inductive inference. Transductive training refers to tasks that have a specific set of test inputs. On the other hand, inductive learning refers to training for a more general problem on unknown inputs. Similar to other NNs, GNNs can be trained in supervised and unsupervised ways.

% (How do you train them; supervised, unsupervised)
%Supervised learning is when the training data is labeled with known outputs so that patterns can be %recognized between the inputs and outputs. With unsupervised training, the training data does not have %labels and the model instead uses a loss heuristic, but these are often challenging to formulate for a %given problem. 

% (Mention the isomorphism issue)
%A significant advantage that GNNs have over CNNs is that they can correctly predict outcomes for isomorphic %graphs. If two graphs have a one-to-one mapping with different node ordering, they are isomorphic. In these %two cases, CNNs will often predict different results for isomorphic graphs depending on training and node %ordering, whereas GNNs predict consistently by aggregating features from topological neighbours. 

%(What are GNNs, how are they different from regular NNs, advantages, disadvantages...)
% this is the basic form of message passing. The gnn survey paper has a more generalized one we can use
The message passing feature is the core principle behind GNNs. This is an iterative process of updating the features (H) of the nodes using learnable weights (W).
%\begin{equation}
%    H = \begin{bmatrix} \overrightarrow{h_1} \\ \overrightarrow{h_2} \\ ... \\ \overrightarrow{h_n} %\end{bmatrix}, 
%    W = \begin{bmatrix} w_1 \\ w_2 \\ ... \\ w_n \end{bmatrix}
%\end{equation}
%based on their neighbors' and their own features. Each GNN layer is a parallel iteration over all nodes %which allows features to be updated with information from neighbors that are one step further away. 
A message-passing layer can be written in matrix form as
\begin{equation}
    H_{k+1}=\sigma(A \times H_{k} \times W_k).
\end{equation}
where $H_{k}$ is matrix of stacked feature vectors for each node at the $k^{th}$ layer; $A$ is the adjacency matrix of the graph; $W_k$ is the learnable component in the $k^{th}$ layer; and $\sigma(\cdot)$ is any non-linear activation function. This message-passing method is used as a way to discover feature embeddings for further use with downstream tasks such as node prediction.  

% What about GNN smoothing if too many iterations of message passing?
% Avoid directly talking about message passing here, since it is in the next paragraph
GNNs can have an over-smoothing issue where node features converge to similar values for deeply layered networks~\cite{oversmoothingsurvey}. This can mostly be addressed by tuning the model's hyperparameters to have appropriate levels, but it has also been examined through adaptive mechanisms~\cite{jknet}.

\subsection{GAP}

% GAP structure
The Generalizable Approximate Graph Partitioning (GAP) framework~\cite{gap-old} is a deep learning-based approach using GNNs to efficiently partition graphs into balanced subsets while minimizing edge cuts. GAP consists of an embedding module and a partitioning module. The embedding module is used to extract topological structure of the input graph by leveraging the input node features and the graph connectivity. It's purpose is to learn a new set of node features that best represent the local connectivity. GAP uses GCN \cite{gcn} and GraphSAGE \cite{graphsage} as the layers inside the embedding module to learn the node representations of the graph. The partitioning module takes the learned node representations as the input and generates the probabilities of each node belonging to partitions $S_{1}, S_{2}, ..., S_{k}$. This module consists of a fully connected layer followed by a softmax activation. 

% GAP loss
GAP uses unsupervised learning and introduces a differentiable loss function that captures the minimum-cut objective of graph partitioning while leveraging back-propagation to optimize the model parameters. The loss function is the expected normalized cut which is defined as:
\begin{equation}
\label{eq:ncut}
Ncut(S_1, S_2, ... S_n) = \sum_{k=1}^{n}{\frac{cut(S_k, \bar{S_k})}{vol(S_k, V)}}
\end{equation}
where $S_k$ is partition $k$ of $n$ partitions on $V$ nodes. The volume ($vol$) of a partition is the degree of all nodes belonging to the partition and the cut is the number of nets with at least one node inside ($S_k$) and one outside ($\bar{S_k}$) of the partition. The $cut$ of one partition $k$ is 
\begin{equation}
cut(S_k,\bar{S_k}) = \sum_{v_i \in S_k, v_j \in \bar{S_k}} e(v_i, v_j)
\end{equation} 
for one partition and half the sum of all partition cuts when considering the cut of an entire graph.

% Why GAP?
The GAP paper~\cite{gap-old} performs an analysis of inductive learning which uses models trained on specific graphs to perform effectively on unseen ones. GAP incorporates graph embedding techniques using PCA to account for graph structures. 

% GAP and balance
GAP indirectly considers partition balance through the volume portion of the loss function. Since the cut size of a partition is normalized to the volume, unequal volumes will indirectly result in unequal cut sizes. Later versions of GAP introduced an explicit balance loss, but this loss is not normalized and is the total difference of partition sizes from the ideal partition size. This means that combining the two loss functions results in primarily optimizing for balanced partitions and not minimum cut.

% How GAP did, why it is incomplete
GAP demonstrated significant improvements of up to 100$\times$ in speed and simultaneous improvements in solution quality compared to hMETIS on \erdos random graphs~\cite{erdos_graphs}, scale free random graphs~\cite{scale_free}, and computation graphs~\cite{gap-old}. However, they did not examine VLSI design graphs which have unique properties as mentioned earlier.

\subsection{Net Models}

% Net model issues
The most significant difference between VLSI design graphs and other typical graphs is that VLSI applications use hypergraphs. In a directed (or undirected) graph, an edge ($e_{ij}$) is a connection between node $n_i$ and $n_j$. In a hypergraph, a hyperedge is typically represented as a set of two or more nodes $e_h = \{n_i, n_j, ...\}$. During graph partitioning, a cut of a hyperedge will result in a cut of one just like a cut of an undirected edge. Often, hypergraphs are decomposed into simpler undirected or directed graphs. 

% I don't think we need to refer to \cite{higher-order-learning} as these net models ahve been around for decades

Clique expansion is one model that enumerates all possible pairs of edges to form a complete graph to represent a hyperedge, so it grows quadratically in size. The decomposed edges are often given partial weight depending on the size of the edge to not change the total edge weight of the graph. While this has complete connectivity information, it is also problematic when considering cut sizes because you cannot distinguish between different hyperedges. 
%For example, if you have a hyperedge, $\{n_1, n_2, n_3, n_4, n_5\}$, you cannot distinguish it from three %hyperedges $\{n_1, n_2, n_3\}$, $\{n_3, n_4, n_5\}$, and $\{n_1, n_2, n_4, n_5\}$, and the cut size could %be incorrect. 

Star expansion is a model that connects all nodes in a hyperedge to an extra zero weight node. The star model requires identifying star nodes separately from design nodes by zero weight so that it does not affect balance, but has the advantage that it grows only linearly in complexity. In partitioning, for example, the star node must also be added to a partition to determine the cut size.

A fanout expansion (or sometimes called source/sink expansion)  is a partial clique model that distinguishes the source or driver node of a hyperedge in a VLSI circuit from the other fanout or sink nodes. This model therefore only enumerates all edges from the driver and grows linearly. 

There has been some work on Hypergraph Neural Networks~\cite{hgnn}, but these have not been extended to GNN partitioning or pooling.

%% file: implementation.tex
\section{Benchmark Implementation}
\label{sec:implementation} 

% What we use
We propose a new set of synthetic benchmarks called \ourNameLong. Our benchmarks consider a Net Distribution Vector (NDV) to describe the fan-out of nets in a design, but we also utilize a Gate Distribution Vector (GDV) which simultaneously describes the number of inputs/outputs on gates in a design. Each of these is a discrete probability distribution function describing the sizes and their given probabilities extracted from real designs allowing the flexibility to rescale the graphs as needed.

% cut size 
% BEKU/MEKU were intended to replicate existing designs but with a ``known upper bound" of the cut size. To be clear, the number of gates/nodes and edges were the same as input benchmarks in BEKU/MEKU. In addition, the cut size was an input to the algorithm ranging form 20-35\% of the edges with no formal justification.  Instead of this, \ourName  uses Rent's rule to determine the number of pins between partitions and, therefore, the number of cut nets. The number of cut nets changes depending on the size of the design according to Rent's rule and the associated constants.

% NDV 
% The NDV used by BEKU/MEKU was an absolute number of nets of each size, just like the benchmark. They did not propose a way to scale the benchmark size.

% GDV
% BEKU/MEKU also did not consider gate fan-in and fan-out which means that the typical number of pins on gates/nodes will diverge from typical designs. Usually, digital logic gates have of 2-6 pins gates with the majority being 2 or 3. \ourName uses the GDV which describes the gates/nodes and their number of pins to select a typical distribution. 
   
% Moreover, the BEKU/MEKU benchmarks are no longer available. The only thing available is the pseudocode in their publication~\cite{beku} which has some ambiguities.

% our algorithm and its inputs
Algorithm~\ref{algo} details the pseudo-code for our proposed \ourName benchmarks. Using the probability distribution functions for NDV and GDV allow us the flexibility to scale the number of nodes in the graph which is an input to the algorithm. We denote a dictionary NDV with key-value pairs $(s,p)$ where $s \ge 2$ is the net-size and $p \in [0,1]$ represent the percentage of this net-size in the design. Similarly, we denote a dictionary GDV with key-value pairs $(g,p)$ where $g \ge 2$ is the gate-pins and $p \in [0,1]$ represent the percentage of this gate in the design. 

% How to use GDV, NDV, and Rent's
We do a weighted sample from the GDV to get the maximum degree ($max\_deg$) of each node which ensures our design follows the GDV distribution. We determine the number of cut nets ($cut\_pins$) using the Rent's rule. Using the total number of components in each partition, the number of pins on that partition will determine the number of cut nets. Our algorithm adds cut nets until the predetermined number is met. Then it continues adding uncut nets until the current degree ($cur\_deg$) of each node reaches the maximum degree ($max\_deg$) limit. We do a weighted sample of the net sizes ($net\_size$) from the NDV which ensures our design follows the NDV distribution.

% Optionally, a number of fixed terminals/nodes can be selected after a synthetic graph is generated to represent fixed input/output pins. The number of fixed terminals can also be derived using Rent's rule for the whole design rather than a single partition.  

%NDV
%GDV
\begin{algorithm}
\DontPrintSemicolon
\SetKwComment{Comment}{/* }{ */}
\caption{\ourNameLong Graph Generation}
\label{algo}
    \SetKwInOut{Input}{Input}\SetKwInOut{Output}{Output}
    \Input{Number of nodes: $n > 0$  \\ Number of partitions: $k \geq 2$ \\ $NDV = \{(s_1, p_1), (s_2, p_2), \ldots, (s_n, p_n)\}$ \\ $GDV = \{(g_1, p_1), (g_2, p_2), \ldots, (g_n, p_n)\}$ \\ Rent parameters: $t = 4$, $p=0.665$}
    \Output{Graph: $G(V,E)$}
    \BlankLine
    \nl Create $n$ nodes and add to $V$\;
    \nl $partitions \gets$ shuffle and split $V$ into $k$ equal sets\;
    \nl $cur\_deg \gets \{(n, 0)$ for $n \in V$\}\;
    \nl $max\_deg \gets \{(n, g)$ for $n \in V$, weighted random $g$ from GDV\}\;
    \nl $cut\_pins \gets t \times (\frac{n}{k})^p$, $net\_count \gets 0$ \;
    \nl Mark free nodes with $cur\_deg < max\_deg$\;
    \nl \While{sufficient free nodes for a net}{
    \nl $net\_size \gets$ weighted random net size from NDV\;
    \nl \eIf{$net\_count < cut\_pins$}{
        \nl $n\_part \gets$ random int $1..net\_size$\;
        }{
        \nl $n\_part \gets net\_size$\;
    }
    \nl $n\_cut \gets net\_size - n\_part$\;

    \nl $part \gets$ randomly pick $part \in partitions$\;
    \nl $o\_part \gets$ $V - part$\;
    \nl $part\_nodes \gets$ random sample $n\_part$ free nodes in $part$, next net if unable\; 
    \nl $cut\_nodes \gets$ random sample $n\_cut$ free nodes in $o\_part$, next net if unable\; 
    \nl $net \gets part\_nodes \cup cut\_nodes$\;
    \nl Add $net$ to $E$, $net\_count+=1$\;
    \nl Increment $cur\_deg$ for all $n \in net$\;
    \nl Update free nodes with $cur\_deg < max\_deg$\;
    }
    % \nl \(Optional\): Mark $f \times n$ nodes as fixed terminals\;
    \nl Run FM to refine partitions
\end{algorithm}

% Comparison with BEKU/MEKU is hard because a) we are making a lot of assumptions about what they did b) according to Rents rule, a constant B is WRONG

% How we extract NDV and GDV from IBM benchmarks (Average of all benchmarks?)

%% file: methodology.tex
\section{Methodology}
\label{sec:methodology}

We implemented our model using the Tensorflow~\cite{tensorflow} and Spektral~\cite{spektral} libraries. We used the same model parameters as in GAP-Random~\cite{gap-old}: the embedding module uses GraphSAGE with 2 layers of 256 units and shared pooling, the partitioning module is 3 dense layers of 128 units with softmax, and the learning rate is $7.5e^{-6}$. We trained the model for 5000 epochs using early stopping with a patience of 5 epochs.

All the graph generation and serial execution was done on a server with two AMD EPYC 7542 32-core  2.9 GHz processors (128 threads total) and 512GiB DRAM. Parallel training, testing, and evaluation are done with an NVIDIA GeForce RTX 4090 24GiB GPU. 

% IBM data & Rent & Erdos graphs
We extracted the total gates, total nets, gate pins of each gate, and net sizes from 8 designs (ibm01, ibm02, ibm07, ibm08, ibm09, ibm10, ibm11, ibm12) in the ISPD98 benchmark suite \cite{ISPD98, ispd98-link}. These benchmarks have unconnected external pins so we remove nets attached to those pins and we also removed diode gates with one pin. We took the mean of all benchmarks to create the NDV and GDV. For the Rent's parameters, we use the values of $t=4$ and $p=0.665$ which are typical for many designs~\cite{rent_properties}. For generating the random \erdos graphs, we used NetworkX~\cite{networkx} with a probability of having an edge between any two nodes being $0.1$ as in GAP~\cite{gap-old}.
%For 1000 node graphs (500 node partitions), for example, this would imply about 250 cut nets in a %bipartitioning since:
%\begin{equation}
%    T=t\times g^p=4 \times 500^{0.665} \approx 250.
%\end{equation}

% How many graphs? Train/dev/test breakdown?  
For training and evaluating the model we used graphs from \ourName. In general, we generated datasets with multiple graphs totalling 100k nodes. We then trained on 90\% of the graphs while examining 10\% for validation. Finally, we tested on 10 random graphs. When training diverges from this, we specify in later sections. For converting the hypergraph to simple graph, by default we use the clique expansion unless otherwise stated.

For evaluating the model's partition quality we use two criteria: cut size and balancedness. The cut size reported is the mean over the test graphs unless otherwise stated. Balancedness is the ratio of nodes in the largest partition over the total nodes and is also reported as the mean over the test graphs. We only report balancedness when necessary, if all the models produce balanced partitions ($\le 5\%$ imbalance) we do not explicitly show it. 

Since the original BEKU/MEKU script and benchmarks are unavailable, we implemented a version of BEKU/MEKU according to the pseudocode from their paper~\cite{beku} for comparison with our graphs.

hMETIS was run with default options and a balance constraint of 5\%. FM was run with a random initial partitioning and  a balance constraint of 5\%.

%% file: results.tex
\section{Evaluation of Benchmarks}
\label{sec:structure}

% How far apart are nodes in a VLSI graph? 
% This might have an affect on the needed number of layers in a GNN.
We first perform an analysis on the structure and topology of the benchmark graphs including the \erdos graphs (used by GAP), the BEKU graphs, and our new \ourName graphs. We generated equal node sized graphs for BEKU and \ourName. However, an equal node size \erdos graph would have over 6.5 million edges due to their highly dense nature, so instead we used a 1000 node \erdos graph which still has significantly different characteristics. 

% Neighborhood and why \erdos are bad
Table~\ref{table:stats} shows the characteristics of the different graphs and it can be clearly seen that the \erdos graphs are densely connected with over $5\times$ more edges and with an average path length being $1.89$. Whereas, for the other benchmarks (and realistic circuits) it's much larger. As can be seen, BEKU and \ourName have similar numbers of edges while \ourName has a longer average path between nodes.

\begin{table}[htb]
\centering
\def\arraystretch{1.5}
\caption{Benchmark Graph Structure Comparison}
\label{table:stats}
\begin{tabular}{|l|c|c|c|}
\hline
                       & \erdos & BEKU & \ourName  \\ \hline
Num. Nodes           & 1000   & 11161    & 11161             \\ \hline     
Num. Edges           & 50020   & 10985    & 10997     \\ \hline
Average path           & 1.89   & 3.49    & 3.69     \\ \hline
%Longest short path  & 3   & 8      & 8     \\ \hline
\end{tabular}
\vspace{-0.3cm}
\end{table}

%\begin{table}[htb]
%\centering
%\def\arraystretch{1.5}
%\caption{Benchmark Graph Structure Comparison}
%\label{table:stats}
%\begin{tabular}{|l|c|c|c|c|}
%\hline
%                       & \erdos & BEKU & \ourName & IBM \\ \hline
%Num. Nodes           & 1000   & 11161    & 11161   &  11161          \\ \hline     
%Num. Edges           & 50020   & 10985    & 10997  &     10956    \\ \hline
%Average path           & 1.89   & 3.49    & 3.69   &    6.56   \\ \hline
%Longest short path  & 3   & 8      & 8    &  15  \\ \hline
%\end{tabular}
%\vspace{-0.3cm}
%\end{table}

%\begin{table}[htb]
%\centering
%\def\arraystretch{1.5}
%\caption{Graph Structure Statistics}
%\label{table:stats}
%\begin{tabular}{|l|c|c|c|c|}
%\hline
%                       & \erdos & BEKU & \ourName  & IBM01\\ \hline
%Num. Nodes           & 1000   & 11201    & 11021 & 11201              \\ \hline     
%Num. Edges           & 50020   & 11055    & 11225 & 11025          \\ \hline
%Average path           & 1.89   & 3.50    & 3.68 &          \\ \hline
%Longest short path  & 3   & 7      & 8  &         \\ \hline
%\end{tabular}
%\vspace{-0.3cm}
%\end{table}

% Why avg and longest path might be an issue
The structure of the graph impacts how well (or poorly) a GNN model receive messages from its neighborhood and learns features that help optimize its objective function. The learnable neighborhood of a node is directly related to its average and furthest distance to other nodes. Each message-passing layer aggregates one-hop-away neighbors. The number of message-passing layers that GAP uses (2) may reduce the effectiveness on these graphs as it will not be able to learn from a wide enough neighborhood to learn about their topology. However, this imposes another challenge: simply increasing the layers is not a solution since GNNs suffer from over-smoothing of features when aggregating over a very large neighborhood. However, the PCA components have a global view of the graph topology which may compensate for this.

\begin{figure}[htb] % t=top, b=bottom, h=here
\begin{center}
\vspace{-0.3cm}
\includegraphics[width = 0.5\textwidth]{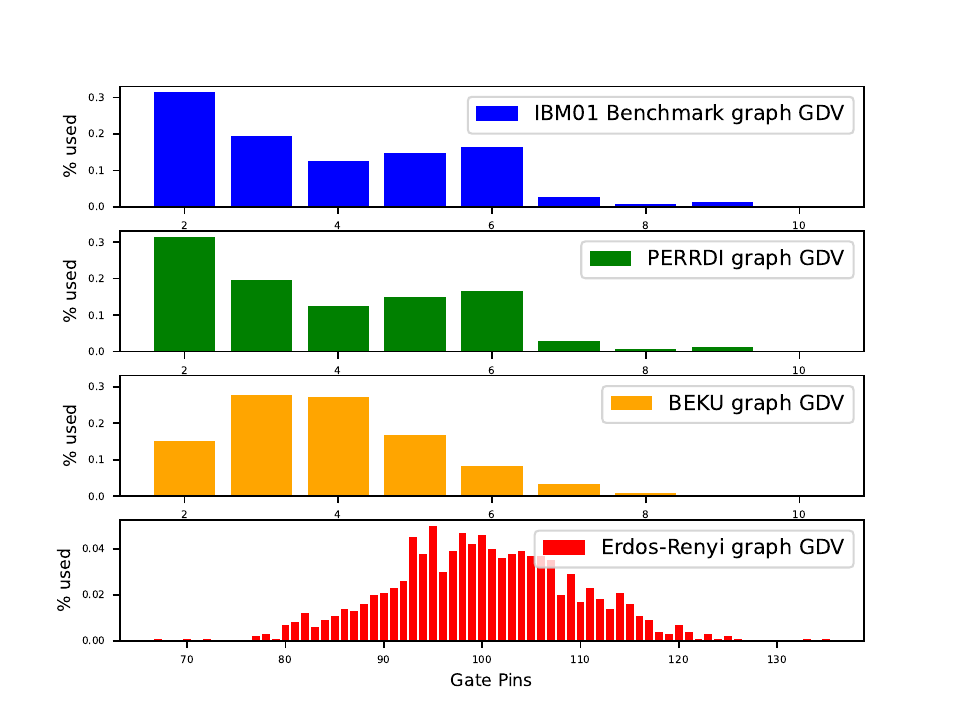}
 \caption { Our proposed benchmark generation follows the GDV of the ISPD benchmarks exactly whereas other graphs were considerably different.
 \label{fig:gdv}}
 \vspace{-0.4cm}
 \end{center}
\end{figure}

% How the GDV differs
We also show GDV plots in Figure~\ref{fig:gdv} since it has a significant contribution to the neighborhood size. We generated a single graph using statistics from the ibm01 benchmark design statistics with BEKU and \ourName and the figure shows that our algorithm matches the benchmark GDV. Although the BEKU algorithm uses similar range of gate pins in the graph, the distribution is significantly different. Specifically, in the BEKU graph, the gates with two pins are not the majority whereas in actual designs and the \ourName graphs, inverters are the most common gate. Similarly, the BEKU graph significantly over-estimates the 3- and 4-pin gates. The \erdos random graphs used by GAP have a completely different distribution with pins ranging from 80-120 pins per node which are unrealistic for VLSI designs.

\section{Evaluation of Partitioning Algorithms}

\subsection{Graph Inference}
\label{sec:inference}

In this section, we evaluate how well the GAP model can infer on new graphs and compare how it does with state-of-the-art partitioner (hMETIS)~\cite{hmetis}, fundamental heuristics (FM), and the known upper bound from \ourName. 

\begin{figure}[htb] % t=top, b=bottom, h=here
\begin{center}
\vspace{-0.4cm}
\includegraphics[width = 0.45\textwidth]{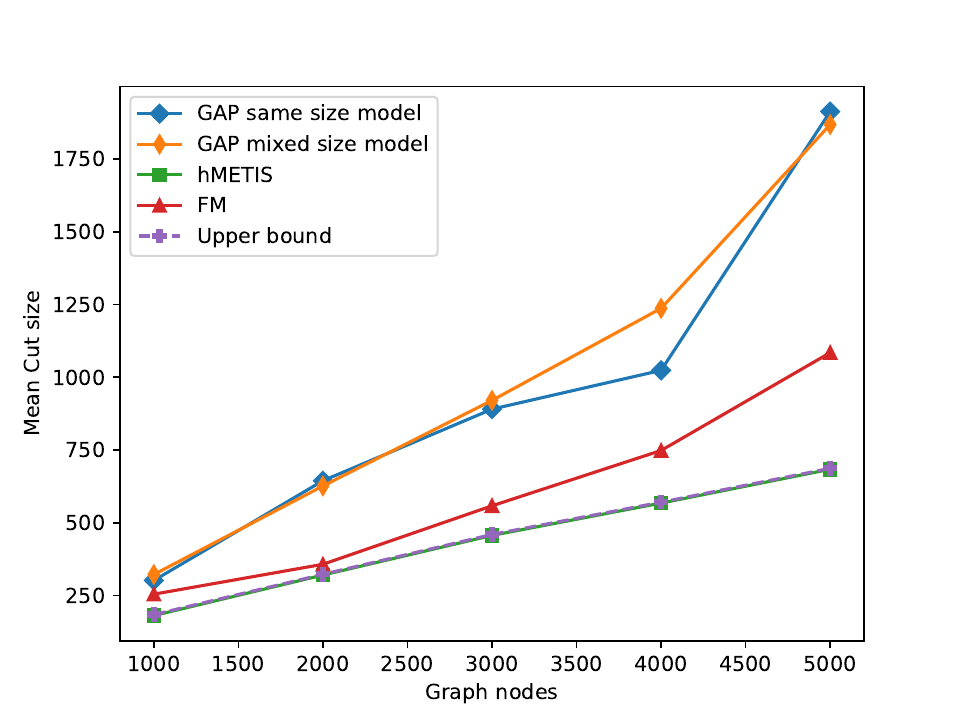}
 \caption {The difference in cut-size  increases between GAP and hMETIS/known upper bound as the problem size grows while GAP does not benefit from size-specific training.
 \label{fig:inductive_learning}}
 \vspace{-0.5cm}
 \end{center}
\end{figure}

In order to study how well GAP performs on VLSI like graphs from \ourName, we generated the training/validation/test set and used GAP alongside hMETIS, FM and Upper Bound from our graphs. Figure~\ref{fig:inductive_learning} shows this comparison.

%1000 nodes 5 graphs, 900 nodes 10 graphs, 800 nodes 15 graphs, 700 nodes 20 graphs, 600 nodes 25 graphs, 500 nodes 60 graphs, 400 nodes 20 graphs, 300 nodes 15 graphs, 200 nodes 10 graphs and 100 nodes 5 graphs)
The ``GAP same size model" is a unique model specifically trained for each graph of a given node size and is considered as the baseline. We further explored how mixing the dataset with different varying graph sizes would affect the learning. For this, we trained ``GAP mixed size model" where we provide a set of varying number and size: $5 \times 100$, $10 \times 200$, $15 \times 300$, $20 \times 400$, $60 \times 500$, $25 \times 600$, $20 \times 700$, $15 \times 800$, $10 \times 900$, and $5 \times 1000$. These had a total of $100,000$ nodes. We used these to train the model and reserved a set to test on the ranges of graphs shown in the Figure~\ref{fig:inductive_learning}. Note that even with graphs of sizes less than $5000$ nodes, the GAP mixed model was able to achieve almost as good results as the baseline. However, apart from that there is a significant difference in terms of cut-size that GAP is able to achieve in comparison with other partitioners.  

\subsection{Run-Time}
\label{sec:runtime}

\begin{figure}[htb] % t=top, b=bottom, h=here
\begin{center}
\vspace{-0.3cm}
\includegraphics[width = 0.45\textwidth]{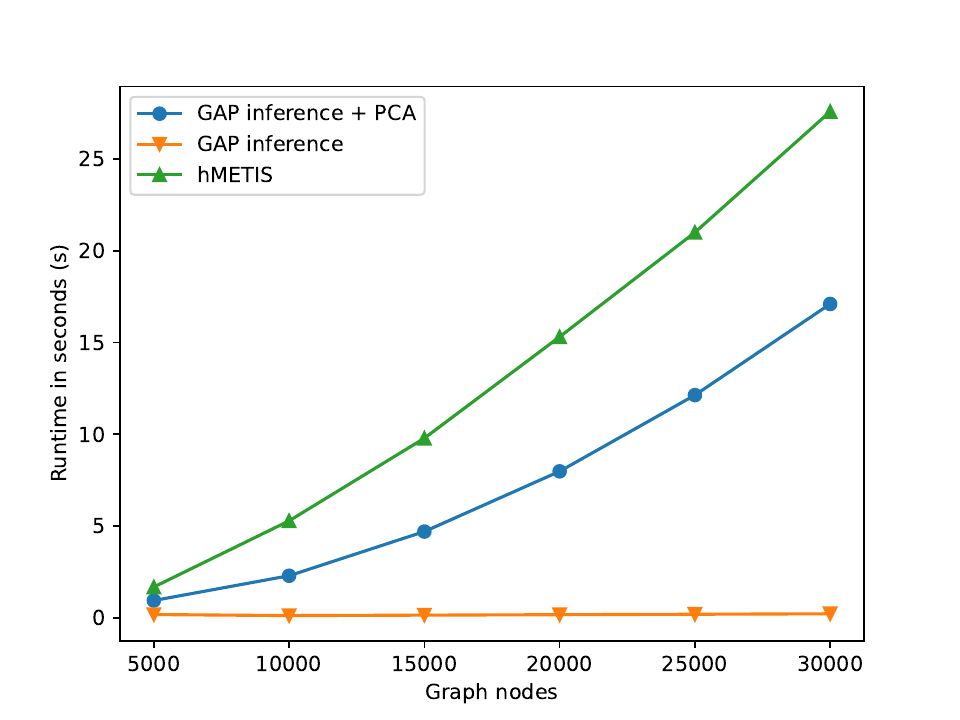}
 \caption {GAP run-time is nearly constant even for moderate graph sizes while PCA feature computation can increase overall run-time but less than hMETIS.
 \label{fig:runtime}}
 \vspace{-0.4cm}
 \end{center}
\end{figure}

The distinct advantage of GAP compared to other partitioning methods is run-time. We evaluated on graphs ranging from $5000$ to $30000$ nodes and show the results in Figure~\ref{fig:runtime}. GAP inference run-time is nearly constant but computation of the PCA components for input features can be significant. We used PyCUDA~\cite{pycuda} to run parallel PCA computation for $1000$ components for all sizes of graphs and included that along with GAP inference runtime. hMETIS, however, has significant increase in run-time compared to both inference and inference including PCA.

The GAP run-time remains constant until GPU memory is exhausted, but GNNs use sparse matrices for the adjacency matrix which enables very large graphs. The feature matrix, however, uses PCA components and is dense, so we discuss the size needed for these features later in Section~\ref{sec:pca}.

\subsection{Net Model}
\label{sec:net_comparison}

\begin{figure}[htb] % t=top, b=bottom, h=here
\begin{center}
\vspace{-0.3cm}
\includegraphics[width = 0.45\textwidth]{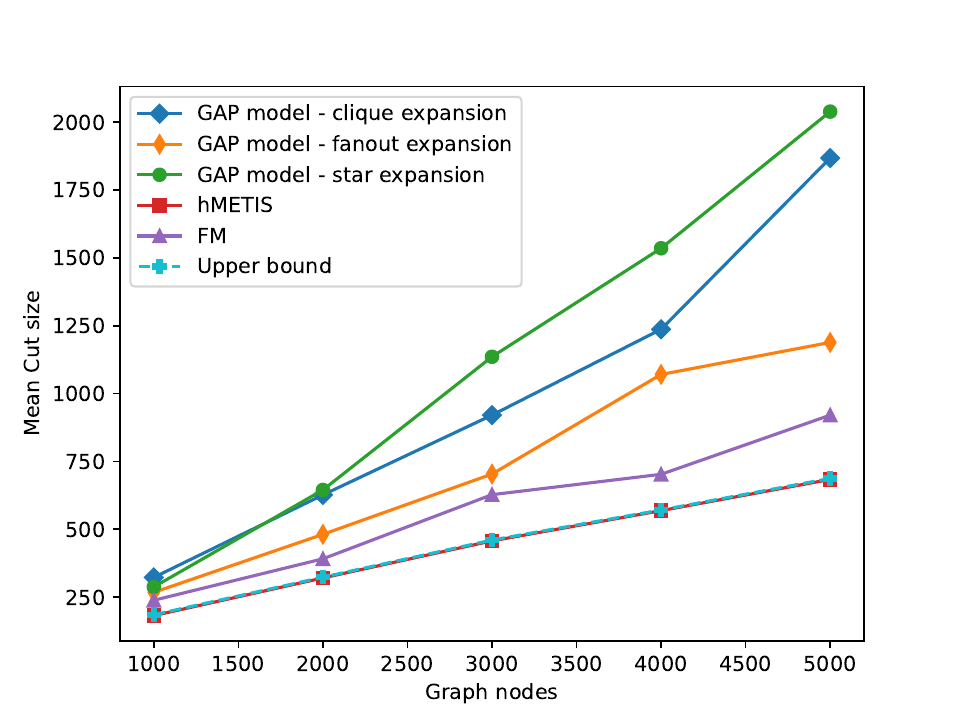}
 \caption {VLSI designs use hypergraphs but the hypergraph net model significantly affects quality of results.
 \label{fig:net-model}}
 \vspace{-0.4cm}
 \end{center}
\end{figure}

% The other challenging with GAP is that it only works on undirected graphs. In particular, PCA to generate the features requires the Laplacian. In addition, GNNs are mostly restricted to undirected graphs.
% Clique vs star vs fan-out model in GAP
GNNs do not inherently support hyperedges so this requires modeling hypergraphs as simple graphs to utilize all of the layers and features available to GNNs.  We use the mixed size training set for this experiment (i.e. one trained model for all GAP experiments) and evaluate on $10$ random graphs of each size. We show the results of three different models (clique, fan-out, and star) and comapre the results with hMETIS, FM, and the known upper bound in Figure~\ref{fig:net-model}. The results show that the fan-out model has a significant improvement over both the clique and star models but there is still sub-optimality compared to the known upper bound and other partitioners.  

\subsection{Balanced Partitioning}
\label{sec:balance}

% Implicit balancing, when doesn't it work?
GAP does not explicitly have a balance constraint, but it is indirectly considered in the normalized expected cut loss function, shown in Equation~\ref{eq:ncut}. The denominator has the volume of a partition which is the total degree of nodes that belong to a partition:
\begin{equation}
    vol(S_k,V) = \sum_{v_i \in S_k, v_j \in V} e(v_i, v_j).
\end{equation}
Since the volume normalizes the cut of each partition relative to the total connections (degree), an imbalance of volume would result in an imbalance of normalized cut. In bi-partitioning, for example, the absolute cut of each partition is the same, but normalization by different volumes would result in an imbalance. While this is not exactly a balance, it has an approximate effect when given a large number of edges and the NDV of those nodes. 

% Why explicit balancing doesn't work
An explicit balance constraint was added in an updated version of GAP~\cite{gap-updated} that uses an additional loss component to quantify the different number of nodes in each partition compared to the expected number of nodes. However, this dominates the cut size loss function in practice resulting in significant degradation of cut size at the expense of highly balanced partitions. 

\subsection{Graph Topology and Embedding}
\label{sec:pca}

The graph structure has a direct impact on a graph's topology. Since GAP uses initial PCA components as features for the embedding, we explore the sensitivity to the number of these components used. The original GAP work used all of the features (e.g., for $1000$ nodes, it used $1000$ features) which would be too many for realistic VLSI designs and would result in slow training times. Also it might over-shadow  other important features one might want to incorporate for the GNNs to learn from such as timing, reliability, and power.

\begin{figure}[htb] % t=top, b=bottom, h=here
\begin{center}
\vspace{-0.3cm}
\includegraphics[width = 0.45\textwidth]{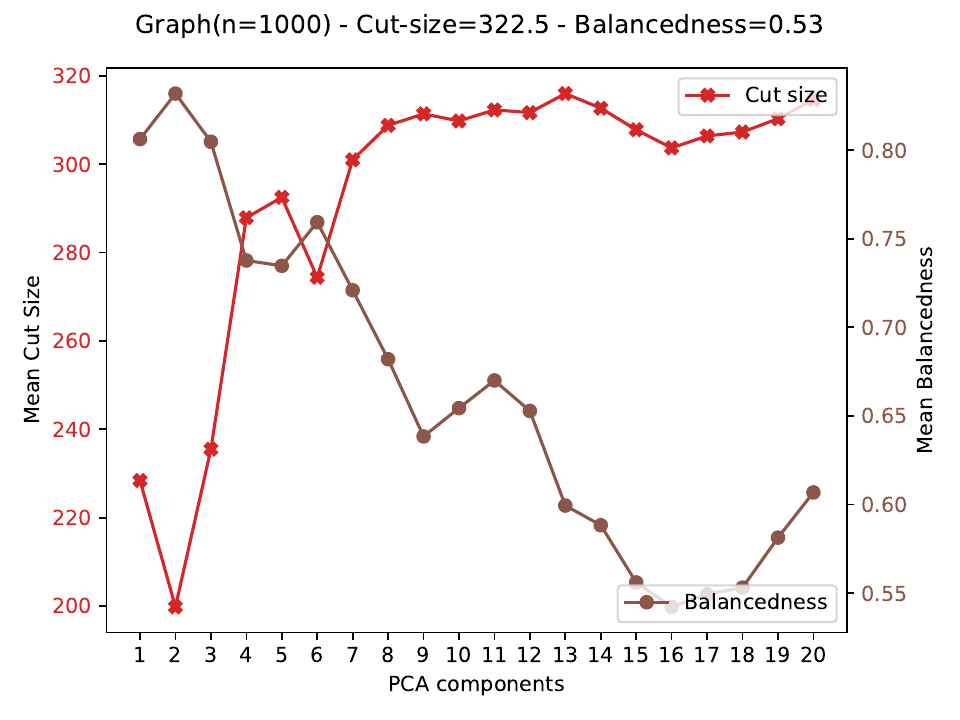}
 \caption {GAP approaches the min cut-size with balance when at least 9 PCA components are used as input features while more does not always improve the cut-size.
 \label{fig:pca}}
 \vspace{-0.4cm}
 \end{center}
\end{figure}

Related to this, previous partitioning works have shown that more Eigenvectors are better for spectral partitioning~\cite{alpert1995spectral}, but we have found that GAP achieves a minimal cut size with far fewer features. We analyzed this by training a model with a mixed-size set of graphs and up to $1000$ features (smaller graphs used fewer features and were padded by 0's).  During inference on $10$ test graphs of $1000$ nodes, we varied the number of features used and the results are shown in Figure~\ref{fig:pca} with a mean known upper bound cut-size of $322.5$.  With lower numbers of features, the balancedness is quite high which can lead to an artificially low cut-size but the upper bound is best with at least $9$ PCA components while balancedness stabilizes.  The small increase of balancedness on the right is not a trend with further increases. 

% Wait for next paper
%\subsection{Fixed Terminals/Nodes}
%GAP does not support fixed terminals while hMETIS does. We are able to add fixed terminals to our synthetic benchmarks simply by selecting some of the nodes as fixed in a partition after graph creation. 

% Wait for next paper
%\subsection{Cell Sizes - maybe}
%\fixme{Can we add sizes of nodes to our benchmarks?}
%VLSI designs have different cell sizes depending on the number of transistors and sizes in a cell to implement a particular logic function and drive strenth. 

\subsection{Multi-way Partitioning}

GAP can perform multi-way partitioning by adjusting the final dense layer to have softmax $k$ outputs for k-way partitioning. We trained and compared with hMETIS using the \ourName synthetic benchmarks of $5000$ nodes. \ourName relies on a bipartitioning FM algorithm for refinement, so we do not do refinement on graphs with more than two partitions. We trained on $10$ graphs and performed validation on $2$ graphs. We performed final testing on $10$ other graphs.

Our results, shown in Table~\ref{table:kway}, show that hMETIS achieves good balance and even beats the known upper bound since it is not refined using multi-way FM. GAP, however, struggles to perform multi-way partitioning. In the case of 8-way partitioning, it leaves 2 partitions completely empty resulting in a balance of $0.409$. For lower 2- and 4-way partitioning, there is a more reasonable balance, but the cut size is dramatically larger than the known upper bound. As demonstrated in Section~\ref{sec:runtime}, the run-time (not shown) of GAP is nearly constant for inference on multi-way graphs.

%\begin{table}[htb]
%\centering
%\def\arraystretch{1.5}
%\caption{Multi-way partitioning on 10000 nodes using GAP has deficiencies compared to hMETIS and the Known %Upper Bound solutions.}
%\label{table:kway}
%\begin{tabular}{|l||r|r|r|r|r|r|}
%\hline
%                         & \multicolumn{2}{|c|}{2-way} & \multicolumn{2}{|c|}{4-way} & \multicolumn{2}{|c|}{8-way} \\ \hline 
%Partitioner             & Cut & Bal & Cut & Bal & Cut & Bal  \\ \hline \hline
%GAP             & 4136 & 0.590 & 5001 & 0.420 & 2769 & 0.606  \\ \hline
%hMETIS             & 1174 & 0.504 & 860 & 0.252 & 571 & 0.126  \\ \hline
%Known UB            & 1176 & 0.505 & 962 & 0.250 & 603 & 0.125  \\ \hline
%\end{tabular}
%\end{table}

\begin{table}[htb]
\centering
\def\arraystretch{1.5}
\caption{Multi-way partitioning on 5000 nodes using GAP has deficiencies compared to hMETIS and the Known Upper Bound solutions.}
\label{table:kway}
\begin{tabular}{|l||r|r|r|r|r|r|}
\hline
                         & \multicolumn{2}{|c|}{2-way} & \multicolumn{2}{|c|}{4-way} & \multicolumn{2}{|c|}{8-way} \\ \hline 
Partitioner             & Cut & Bal & Cut & Bal & Cut & Bal  \\ \hline \hline
GAP             & 1573 & 0.591 & 1659 & 0.332 & 903 & 0.409  \\ \hline
hMETIS             & 684 & 0.506 & 516 & 0.251 & 354 & 0.126  \\ \hline
Known UB            & 685 & 0.509 & 594 & 0.250 & 378 & 0.125  \\ \hline
\end{tabular}
\vspace{-0.3cm}
\end{table}

%\begin{table}[htb]
%\centering
%\def\arraystretch{1.5}
%\caption{Multi-way partitioning on 2000 nodes using GAP with new model parameters.}
%\label{table:kway}
%\begin{tabular}{|l||r|r|r|r|r|r|}
%\hline
%                         & \multicolumn{2}{|c|}{2-way} & \multicolumn{2}{|c|}{4-way} & \multicolumn{2}{|c|}{8-way} \\ \hline 
%Partitioner             & Cut & Bal & Cut & Bal & Cut & Bal  \\ \hline \hline
%GAP             & 501 & 0.551 & 418 & 0.780 & 432 & 0.491  \\ \hline
%hMETIS             & 329 & 0.520 & 266 & 0.255 & 189 & 0.127  \\ \hline
%Known UB            & 332 & 0.514 & 317 & 0.250 & 206 & 0.125  \\ \hline
%\end{tabular}
%\vspace{-0.3cm}
%\end{table}

%% file: conclusion.tex
\section{Conclusion}

In this paper, we explored the advantages and disadvantages of applying Deep Learning (DL) to VLSI design partitioning. We introduced and examined a new set of synthetic benchmarks, \ourName, with known upper bound solutions and VLSI design-like characteristics. Our study demonstrated DL models' capability to infer partitions on novel, unseen VLSI design graphs, highlighting trade-offs with training methodologies, net models, and number of input features. Notably, our findings show significant potential for run-time improvement with DL-based partitioning, but we also identify areas for enhancement, including cut-size quality, balance constraints, and k-way partitioning extensions.

%% file: main.bbl
\begin{thebibliography}{10}

\bibitem{fm}
C.~M. Fiduccia and R.~M. Mattheyses, ``A linear-time heuristic for improving network partitions,'' {\em Papers on Twenty-five years of electronic design automation}, pp.~241--247, 1988.

\bibitem{kl}
B.~W. Kernighan and S.~Lin, ``An efficient heuristic procedure for partitioning graphs,'' {\em The Bell system technical journal}, vol.~49, no.~2, pp.~291--307, 1970.

\bibitem{spectral}
F.~McSherry, ``Spectral partitioning of random graphs,'' in {\em Proceedings 42nd IEEE Symposium on Foundations of Computer Science}, pp.~529--537, 2001.

\bibitem{mlpart}
A.~E. Caldwell, A.~B. Kahng, and I.~L. Markov, ``Improved algorithms for hypergraph bipartitioning,'' in {\em Proceedings of the 2000 Asia and South Pacific Design Automation Conference}, ASP-DAC '00, (New York, NY, USA), p.~661–666, Association for Computing Machinery, 2000.

\bibitem{hmetis}
G.~Karypis, ``{hMETIS} 1.5 : A hypergraph partitioning package,'' {\em http://www.cs.umn.edu/~metis}, 1998.

\bibitem{gnn_survey}
Y.-C. Lu and S.~K. Lim, ``On advancing physical design using graph neural networks (invited paper),'' in {\em 2022 IEEE/ACM ICCAD}, pp.~1--7, 2022.

\bibitem{asymcheegercutpool}
J.~B. Hansen and F.~M. Bianchi, ``Total variation graph neural networks,'' 2023.

\bibitem{mincutpool}
F.~M. Bianchi, D.~Grattarola, and C.~Alippi, ``Mincut pooling in graph neural networks,'' {\em CoRR}, vol.~abs/1907.00481, 2019.

\bibitem{dmonpool}
A.~Tsitsulin, J.~Palowitch, B.~Perozzi, and E.~M{\"{u}}ller, ``Graph clustering with graph neural networks,'' {\em CoRR}, vol.~abs/2006.16904, 2020.

\bibitem{diffpool}
R.~Ying, J.~You, C.~Morris, X.~Ren, W.~L. Hamilton, and J.~Leskovec, ``Hierarchical graph representation learning with differentiable pooling,'' {\em CoRR}, vol.~abs/1806.08804, 2018.

\bibitem{gap-old}
A.~Nazi, W.~Hang, A.~Goldie, S.~Ravi, and A.~Mirhoseini, ``{GAP:} generalizable approximate graph partitioning framework,'' {\em CoRR}, vol.~abs/1903.00614, 2019.

\bibitem{alpert1995spectral}
C.~J. Alpert and S.-Z. Yao, ``Spectral partitioning: The more eigenvectors, the better,'' in {\em Proceedings of the 32nd annual ACM/IEEE design automation conference}, pp.~195--200, 1995.

\bibitem{KKMMN2016}
K.~Kersting, N.~M. Kriege, C.~Morris, P.~Mutzel, and M.~Neumann, ``Benchmark data sets for graph kernels,'' 2016.
\newblock \url{http://graphkernels.cs.tu-dortmund.de}.

\bibitem{beku}
J.~Cong, M.~Romesis, and M.~Xie, ``Optimality, scalability and stability study of partitioning and placement algorithms,'' in {\em Proceedings of the 2003 International Symposium on Physical Design}, ISPD '03, (New York, NY, USA), p.~88–94, Association for Computing Machinery, 2003.

\bibitem{cong}
J.~Cong, ``Personal communication,'' 2024.

\bibitem{rent}
M.~Y. Lanzerotti, G.~Fiorenza, and R.~A. Rand, ``Microminiature packaging and integrated circuitry: The work of e. f. rent, with an application to on-chip interconnection requirements,'' {\em IBM Journal of Research and Development}, vol.~49, no.~4.5, pp.~777--803, 2005.

\bibitem{gcn}
T.~N. Kipf and M.~Welling, ``Semi-supervised classification with graph convolutional networks,'' {\em arXiv:1609.02907}, 2017.

\bibitem{gat}
P.~Veličković, G.~Cucurull, A.~Casanova, A.~Romero, P.~Liò, and Y.~Bengio, ``{Graph Attention Networks},'' {\em arXiv:1710.10903}, 2018.

\bibitem{graphsage}
W.~L. Hamilton, R.~Ying, and J.~Leskovec, ``Inductive representation learning on large graphs,'' {\em arXiv:1706.02216}, 2018.

\bibitem{oversmoothingsurvey}
T.~K. Rusch, M.~M. Bronstein, and S.~Mishra, ``A survey on oversmoothing in graph neural networks,'' {\em arXiv:2303.10993}, 2023.

\bibitem{jknet}
K.~Xu, C.~Li, Y.~Tian, T.~Sonobe, K.~ichi Kawarabayashi, and S.~Jegelka, ``Representation learning on graphs with jumping knowledge networks,'' {\em arXiv:1806.03536}, 2018.

\bibitem{erdos_graphs}
P.~Erdös and A.~Rényi, {\em On the evolution of random graphs}, pp.~38--82.
\newblock Princeton: Princeton University Press, 2006.

\bibitem{scale_free}
B.~Bollob\'{a}s, C.~Borgs, J.~Chayes, and O.~Riordan, ``Directed scale-free graphs,'' in {\em Proceedings of the Fourteenth Annual ACM-SIAM Symposium on Discrete Algorithms}, SODA '03, (USA), p.~132–139, Society for Industrial and Applied Mathematics, 2003.

\bibitem{hgnn}
Y.~Feng, H.~You, Z.~Zhang, R.~Ji, and Y.~Gao, ``Hypergraph neural networks,'' {\em CoRR}, vol.~abs/1809.09401, 2018.

\bibitem{tensorflow}
M.~Abadi, A.~Agarwal, P.~Barham, E.~Brevdo, Z.~Chen, C.~Citro, G.~S. Corrado, A.~Davis, J.~Dean, M.~Devin, S.~Ghemawat, I.~Goodfellow, A.~Harp, G.~Irving, M.~Isard, Y.~Jia, R.~Jozefowicz, L.~Kaiser, M.~Kudlur, J.~Levenberg, D.~Man\'{e}, R.~Monga, S.~Moore, D.~Murray, C.~Olah, M.~Schuster, J.~Shlens, B.~Steiner, I.~Sutskever, K.~Talwar, P.~Tucker, V.~Vanhoucke, V.~Vasudevan, F.~Vi\'{e}gas, O.~Vinyals, P.~Warden, M.~Wattenberg, M.~Wicke, Y.~Yu, and X.~Zheng, ``{TensorFlow}: Large-scale machine learning on heterogeneous systems,'' 2015.
\newblock Software available from tensorflow.org.

\bibitem{spektral}
D.~Grattarola and C.~Alippi, ``{Graph Neural Networks in TensorFlow and Keras with Spektral},'' {\em arXiv:2006.12138}, 2020.

\bibitem{ISPD98}
C.~J. Alpert, ``The ispd98 circuit benchmark suite,'' in {\em Proceedings of the 1998 International Symposium on Physical Design}, ISPD '98, (New York, NY, USA), p.~80–85, Association for Computing Machinery, 1998.

\bibitem{ispd98-link}
``The ispd98 benchmark suite,'' 1998.
\newblock https://vlsicad.eecs.umich.edu/.

\bibitem{rent_properties}
P.~Verplaetse, J.~Dambre, D.~Stroobandt, and J.~Van~Campenhout, ``On partitioning vs. placement rent properties,'' in {\em Proceedings of the 2001 International Workshop on System-Level Interconnect Prediction}, SLIP '01, (New York, NY, USA), p.~33–40, Association for Computing Machinery, 2001.

\bibitem{networkx}
A.~Hagberg, P.~Swart, and D.~Chult, ``Exploring network structure, dynamics, and function using networkx,'' in {\em Proceedings of the 7th Python in Science Conference}, 01 2008.

\bibitem{pycuda}
``{PyCUDA},'' {\em https://pypi.org/project/pycuda/}, 2024.

\bibitem{gap-updated}
A.~Nazi, W.~Hang, A.~Goldie, S.~Ravi, and A.~Mirhoseini, ``A deep learning framework for graph partitioning,'' {\em ICLR Workshop}, 2019.

\end{thebibliography}
